\documentclass[12pt,preprint]{aastex}
\usepackage{colordvi}
\usepackage{color}
\begin{document}

\title{On the Relationship Between Coronal Magnetic Decay Index and CME
Speed}

\author{Yan Xu\altaffilmark{1},
        Chang Liu\altaffilmark{1},
        Ju Jing\altaffilmark{1},
        and Haimin Wang\altaffilmark{1}}

\affil{1.\ Space Weather Research Lab, Center for Solar-Terrestial
Research, \\ New Jersey Institute of Technology\\
323 Martin Luther King Blvd, Newark, NJ 07102-1982}

\email{yx2@njit.edu}

\date{\today}

\begin{abstract}

Numerical simulations suggest that kink and torus instabilities
are two potential contributors to the initiation and prorogation
of eruptive events. A magnetic parameter named decay index (i.e.,
the coronal magnetic gradient of the overlying fields above the
eruptive flux ropes) could play an important role in controlling
kinematics of eruptions. Previous studies have identified a
threshold range of the decay index that distinguishes between
eruptive and confined configurations. Here we advance the study by
investigating if there is a clear correlation between the decay
index and CME speed. 38 CMEs associated with filament eruptions
and/or two-ribbon flares are selected using the H$\alpha$ data
from the Global H$\alpha$ Network. The filaments and flare ribbons
observed in H$\alpha$ associated with the CMEs help to locate the
magnetic polarity inversion line, along which the decay index is
calculated based on the potential field extrapolation using MDI
magnetograms as boundary conditions. The speeds of CMEs are
obtained from the LASCO C2 CME catalog available online. We find
that the mean decay index increases with CME speed for those CMEs
with a speed below 1000 km/s, and stays flat around 2.2 for the
CMEs with higher speeds. In addition, we present a case study of a
partial filament eruption, in which the decay indexes show
different values above the erupted/non-erupted part.

\end{abstract}


\section{Introduction}

A coronal mass ejection (CME) is a sudden large-scale eruption
from the solar surface to the interplanetary space. The total mass
carried by a CME could reach the order of $10^{13}$~kg
\citep{Chen2001} and the observed speed is as high as 3387 km/s
according to the online catalog by the Large Angle and
Spectrometric Coronagraph (LASCO). The outward ejections contain
not only massive plasmas but also magnetic fields. An
earth-oriented CME can disrupt the Earth's magnetosphere or even
as deep as the Earth's surface.

Statistical studies \citep{Moon2002, Jing2004} have found that
CMEs are often associated with other phenomena of solar activity,
such as filament eruptions and solar flares. From a theoretical
aspect, they are now believed to be different manifestations of a
single eruptive event \citep[e.g.,][]{Priest2002}, in which
magnetic reconnection is a key process. In an equilibrium state, a
filament may be supported by the magnetic flux rope
\citep{Antiochos1999, Fan2007,Torok2005}. However, the kink and
torus instabilities, determined by the core field of magnetic flux
rope and overlaying field (named strapping field), could lead
filaments to erupt. The kink instability is measured by the number
of magnetic twist. When the twist exceeds a certain threshold, for
instance $2.5~\pi$ to $3.5~\pi$ \citep{Hood1981, Fan2003,
Torok2004}, the filament would become unstable and eruptive. On
the other hand, filament eruptions can also be triggered by torus
instability, as a result of the broken of Shafranov equilibrium
when the field strength decreases faster
\citep{Kliem2006,Torok2005}. In this respect, knowledge of
gradient of the strapping field is crucial for understanding the
physical mechanisms of solar eruptive events.

To quantitatively describe how fast the strapping field decays, a
decay index is defined as: $n = -dlog(B_{t})/dlog(h)$
\citep{Kliem2006}, in which $B_{t}$ is the strength of the
strapping field in transverse direction and $h$ is the radial
height above the photosphere. According to this definition, the
larger value of decay index indicates that the overlying field
strength decreases faster. In practice, the field strengths at
different heights can be obtained by extrapolating coronal field
under the assumption of potential field. A typical range of decay
index $n$ falls into $1.1 \sim 2.0$
\citep{Bateman1978,Demoulin2010,Fan2007,Schrijver2008}. Based on a
sample of ten events, \citet{Liu2008a} find that ``for a failed
eruption, the field strength is about a factor of 3 stronger than
that for a full eruption''. Decay index is usually measured along
the polarity inversion line (PIL). In statistical studies such as
\citet{Liu2008a}, all the decay indexes along a certain PIL were
averaged to yield one single value representing the magnetic
configuration of the corresponding active region. However, in some
particular events, the decay index varies significantly along the
PIL. \citet{Liu2010a} analyzed an eruptive event on 2005 Jan 15,
in which a filament erupted partially. They found that the decay
index above the region of failed filament eruption are much
smaller than those above eruptive regions. More recently,
\citet{Guo2010} analyzed another failed event on 2005 May 27 and
further confirmed that the kink instability initiated this failed
event, but confined by above strapping fields.

The main objective of this paper is to present the statistical
study of correlation between decay index and CME speed. We
collected 38 events and calculated the decay index of each event.
A positive correlation between the CME speed and the decay index
is found. In addition, we present analysis of a partial eruption
on 2000 Sep 20 observed by Big Bear Solar Observatory (BBSO).
Decay indexes along the PIL are derived using Michelson Doppler
Imager (MDI) magnetograms and the different decay indexes above
erupte/non-erupted region were found.

\section{Observations and Measurements}

In order to calculate the decay index, extrapolation of magnetic
field are often used as coronal field measurement is not
available. Non-linear force free field (NLFFF) extrapolation of
high-quality vector magnetograms is believed to represent the 3D
coronal magnetic field accurately. On the other hand, it is
usually reasonable to assume that the strapping field may be more
objectively described by potential fields, as fields turn
potential above certain height even in complicated active regions
\citep{Jing2008}. In addition, NLFFF method requires vector
magnetograms with large field-of-view (FOV), which are unavailable
prior to the launch of Helioseismic and Magnetic Imager on the
Solar and Heliospheric Observatory (SDO/HMI). Therefore, the usage
of potential field extrapolation that only requires the
line-of-sight magnetograms, will expand the number of feasible
events tremendously. Previous studies also suggest that the
coronal field reconstructed under the NLFFF assumption may
approach a potential configuration above an altitude of about 30
Mm \citep[e.g.,][]{Jing2008}. In this study, potential field is
computed using a Green's function method \citep{Chiu1977,
Metcalf2008}. The extrapolation was constructed by 100 grids in
the vertical direction above the photosphere. The grid space is
one pixel. With the MDI's 1.98$\arcsec$ pixel resolution, we are
able to construct a 3D magnetic field with a vertical dimension of
about 140 Mm. In previous studies of \citet{Liu2008a} and
\citet{Liu2010a}, the strapping fields were believed to dominate
in a range of 42 $\sim$ 105 Mm. Xu et al., 2010 also found that a
filament eruption mainly starts from 20 to 60 Mm using the
stereoscopy method with EUV and H$\alpha$ data. Therefore, in this
study, decay indexes are calculated in the same height range of 42
$\sim$ 105 Mm as in the previous studies.

It is well known that the filaments or flux ropes reside above the
PILs which divide opposite magnetic polarity fluxes
\citep{Liu2008a}. Therefore, the decay indexes along the PIL in
principle represent the majority of decay of strapping field and
were commonly calculated in previous analysis \citep{Liu2008a,
Guo2010}. Typically, a PIL is identified by choosing the points
with zero Gauss but large gradient on the magnetograms. However,
in most of events, the PIL extends far away from the center of an
active region or filament of interest, leading to substantial
uncertainties in the averaged decay index along the entire PIL.
Moreover, there are some sections of PIL belong to ``quiet''
regions that do not contribute to the initiation of eruption.
Therefore, we choose to use the extend of filament as a constraint
to define an area within which the PIL is determined. The Sobel
function (an embedded function of IDL) is then applied to the
select area of the magnetogram and find PIL by looking for the
pixels with the highest gradient and near zero field strength.

\begin{figure}
\centering
\includegraphics[scale=1.0]{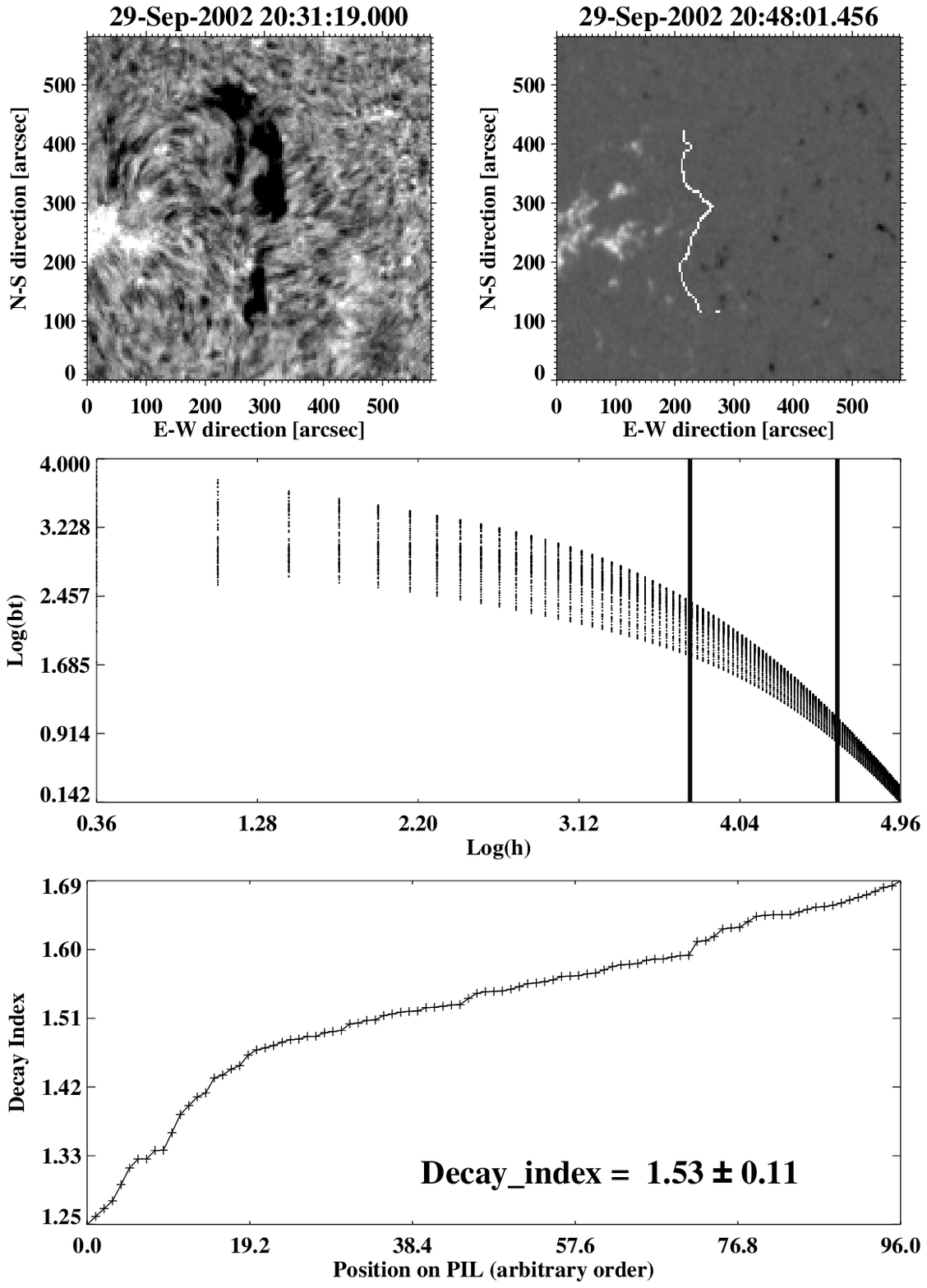}
\caption{Top left panel: An H$\alpha$ image showing the filament
prior to its eruption observed on 2002 September 29. Top right
panel: MDI magnetogram with PIL overploted (white curve). Middle
panel: $Log(B_{t})$ vs. $Log(h)$ along the PIL. The height range
within which the decay index is calculated is marked by two
vertical lines. Bottom panel: Derived decay indexes along the PIL
in an ascending order. In average, the decay index for this event
is 1.53. \label{20020929}}
\end{figure}

As an example, we first present the process of calculating the
decay index of the eruptive filament/CME event on 2002 September
29. This filament erupted around 22:20 UT and the corresponding
CMEs was observed by LASCO at 23:54 UT. Graphical representations
of the details are illustrated in Figure~\ref{20020929}. The top
left shows an H$\alpha$ image of the filament prior to its
eruption at 20:31 UT obtained from the Global H$\alpha$ Network
(GHN). The corresponding MDI magnetogram with the same FOV at
20:48 UT is shown in the top right panel. The PIL under the
eruptive filament is plotted in white color on the top right
panel. In the middle panel, we plot $Log(B_{t})~vs.~ Log(h)$ for
each pixel on the PIL. As mentioned before, the height range of 42
to 105 Mm was used from extrapolated fields to calculate decay
index. In logarithm space, these values are converted to 3.73 and
4.65 as marked by the two vertical lines in the middle panel.
Within such a height range, the derived decay indexes are plotted
in the bottom panel in an ascending order of their values but
regardless their positions on the PIL. The lowest and highest
values of the decay indexes are 1.31 and 1.75, respectively. The
mean decay index with one sigma uncertainty is $1.53 \pm 0.11$.
Note that this is an event occurred in quiet sun. Not surprisingly
we will see in the next section that this decay index is
relatively small comparing with those of active regions.

\section{Statistical Study of Relationship Between Decay Indexes and CME Speed}

Motivated by the above analysis and previous studies
\citep[e.g.,][]{Liu2008a, Liu2010a}, in this paper, we choose to
compare the decay index with the eruption kinematics, in
particular the linear speed of CMEs. As listed in
Table~\ref{list}, 38 events are selected for analysis. To minimize
the projection effect on magnetic field extrapolation, only events
not close to the solar limb (less than 60 degrees from the disk
center in both E-W and N-S directions) were selected. Note that
not all of these events are associated with filament eruptions.
According to \citet{Jing2004}, some CMEs were associated with
flares but no filaments were involved in the eruptions. In some
cases, the absence of filament eruption might be simply due to the
data gap, or because that the material confined in filament
channels is not dense enough to be observed as filaments
\citep{Martin2000}. In such cases, the magnetic configuration is
not necessary to be different from that of events with filament
eruptions. On the other hand, a number of eruptive filaments
located in the quiet Sun do not incur detectable flares in X-ray
emission. For those events without clearly observed filaments in
H$\alpha$ or EUV 304, the flare ribbons are used to outline the
center of energy release area and the PIL is confined in such an
area. In Table~\ref{list}\footnote{Some events were not associated
with detectable flare emission in X-ray. The listed ``flare time''
is actually the time for H$\alpha$ emission or filament
eruption.}, decay indexes of 38 events are listed, as well as the
linear CME velocities obtained from LASCO C2 catalog
(\url{http://cdaw.gsfc.nasa.gov/CME\_list/index.html}). For each
event, the decay index is averaged in the height range of 42-105
Mm. As we can see in Figure 1, normally the decay index does not
change too much within such a height range and the averaged value
is pretty reasonable to present the magnetic gradient along the
PIL.

The data points of these 38 events are presented in
Figure~\ref{f1} and fitted with a third-order polynomial function
($y = a + b x + c x^{2} + d x^{3}$). Two distinct trends are found
of the CME speed as a function of the decay index: 1) Below about
1000~km/s, the CME speed increases with decay index monotonically;
2) For CMEs with higher speed above 1000~km/s, the decay indexes
are almost constant of 2.2. This upper limit is consistent with
theoretical prediction of 2.0 \citep{Kliem2006} and the result of
2.25 from observational study \citep{Liu2008}. In five events, the
erupted filaments were lied above quiet Sun regions and their
decay indexes are among the smallest from 0.98 to 1.68. In
Figure~\ref{f1}, these events are plotted with triangles. We are
not able to draw any statistical conclusion using only five
points, but these events appear to have relative small decay
indexes indicating that the strapping fields of the quiet regions
are relatively strong compared to the initiation force of the
eruption.

\begin{table}[pht]
\caption{Overview of eruption events. \label{list}}

\begin{tabular}{lccccccr}
\\
\tableline\tableline
Date & Flare & Location & Type & CME  & CME   & Decay & Flare   \\
     & Time  &          &      & Time & Speed & Index & Magnitude \\
2000-Jul-14 & 10:10:00 & N18E00 & AR9077 & 10:54:07 & 1674 & 1.98
$\pm$ 0.19 &
X5.7   \\
2000-Sep-12 & 13:00:00 & S27W06 & AR9163 & 11:54:05 & 1550 & 1.64
$\pm$ 0.14 &
M1.0   \\
2000-Sep-16 & 05:05:00 & N14W13 & AR9165 & 05:18:14 & 1215 & 2.17
$\pm$ 0.18 &
M2.1   \\
2000-Nov-24 & 04:55:00 & N21W07 & AR9236 & 05:30:05 & 1289 & 2.67
$\pm$ 0.31 & X2.0 \\
2000-Nov-25 & 00:59:00 & N09E32 & AR9240 & 01:31:58 & 2519 & 2.62
$\pm$ 0.16 &
M8.2   \\
2001-Apr-01 & 11:00:00 & N17W57 & AR9393 & 11:26:06 & 1475 & 2.27
$\pm$ 0.19 &
M2.1   \\
2001-Apr-10 & 05:06:00 & S22W20 & AR9415 & 05:30:00 & 2411 & 2.04
$\pm$ 0.14 &
X2.3   \\
2001-Jun-15 & 10:00:00 & S26E41 & AR9502 & 10:31:33 & 1090 & 2.36
$\pm$ 0.20 &
M6.3   \\
2001-Sep-24 & 09:32:00 & S18E18 & AR9632 & 10:30:59 & 2402 & 2.34
$\pm$ 0.14 &
X2.6   \\
2001-Oct-09 & 10:48:00 & S26E03 & AR9653 & 11:30:00 & 973 & 2.3
$\pm$ 0.17 &
M1.4   \\
2001-Oct-19 & 16:15:00 & N18W40 & AR9661 & 16:50:00 & 901 & 1.93
$\pm$ 0.09 &
X1.6   \\
2001-Oct-22 & 14:27:00 & S19E13 & AR9672 & 15:06:05 & 1366 & 2.2
$\pm$ 0.08 &
M6.7   \\
2001-Nov-04 & 16:03:00 & N05W29 & AR9684 & 16:35:06 & 1810 & 2.2
$\pm$ 0.07 &
X1.0   \\
2002-Jan-28 & 10:00:00 & S23E15 & QS & 10:54:00 & 524 & 1.5 $\pm$
0.08 & C4.6
 \\
2002-Apr-17 & 07:46:00 & S15W42 & AR9906 & 08:26:05 & 1240 & 2.16
$\pm$ 0.08 &
M2.6   \\
2002-Apr-22 & 22:37:00 & S10W03 & QS & 00:38:00 & 286 & 1.4 $\pm$
0.07 & ---
 \\
2002-May-21 & 21:20:00 & N17E38 & AR9960 & 21:50:05 & 853 & 1.85
$\pm$ 0.12 &
M1.5   \\
2002-May-22 & 03:30:00 & S12W60 & QS & 03:50:05 & 1557 & 1.68
$\pm$ 0.09 & C5.0
 \\
2002-Jul-07 & 17:00:00 & N08W49 & QS & 18:06:00 & 750 & 1.36 $\pm$
0.10 & ---
 \\
2002-Jul-29 & 07:13:00 & N32W40 & QS & 11:09:09 & 334 & 0.98 $\pm$
0.10 & ---
 \\
2002-Sep-29 & 22:20:00 & S13E21 & QS & 23:54:00 & 254 & 1.53 $\pm$
0.11 & ---
 \\
2002-Nov-09 & 13:08:00 & S10W42 & AR10180 & 13:31:45 & 1838 & 2.44
$\pm$ 0.18 &
M4.6   \\
2003-May-29 & 00:51:00 & S07W46 & AR10365 & 01:27:12 & 1237 & 2.32
$\pm$ 0.19 &
X1.2   \\
2003-Jun-15 & 22:35:00 & S16W39 & AR10380 & 23:54:05 & 2053 & 2.39
$\pm$ 0.11 &
C2.1   \\
2003-Oct-28 & 09:51:00 & S16E04 & AR10486 & 11:30:05 & 2459 & 2.7
$\pm$ 0.18 &
X17   \\
2003-Oct-29 & 20:40:00 & S17W10 & AR10486 & 20:54:05 & 2029 & 2.55
$\pm$ 0.08 &
X10   \\
2003-Nov-07 & 15:42:00 & N09W08 & AR10696 & 15:54:05 & 2237 & 2.34
$\pm$ 0.09 &
X2.0   \\
2003-Nov-18 & 07:23:00 & N03E08 & AR10501 & 08:06:05 & 1223 & 2.2
$\pm$ 0.21 &
M3.2   \\
2004-Jul-25 & 14:19:00 & N08W35 & AR10653 & 14:54:05 & 1333 & 2.17
$\pm$ 0.17 &
M1.1   \\
2004-Nov-10 & 01:59:00 & N09W49 & AR0696 & 02:26:05 & 3387 & 2.2
$\pm$ 0.14 &
X2.5   \\
2005-Jan-15 & 22:16:00 & N13W04 & AR10720 & 23:06:50 & 2861 & 2.34
$\pm$ 0.10 &
X1.6   \\
2005-Jan-17 & 06:59:00 & N13W29 & AR10720 & 09:54:05 & 2547 & 2.29
$\pm$ 0.14 &
X3.8   \\
2005-May-06 & 16:03:00 & S06E23 & AR10758 & 17:28:00 & 1128 & 1.85
$\pm$ 0.08 &
C8.5   \\
2005-Jul-09 & 21:47:00 & N12W31 & AR10786 & 22:30:05 & 1540 & 2.45
$\pm$ 0.21 &
M2.8   \\
2005-Aug-28 & 10:28:00 & N11E26 & AR10803 & 10:56:07 & 1047 & 2.18
$\pm$ 0.07 &
---   \\
2005-Sep-13 & 19:19:00 & S11E17 & AR10808 & 20:00:05 & 1866 & 1.82
$\pm$ 0.07 &
X1.5   \\
2006-Dec-13 & 02:14:00 & S06W35 & AR10930 & 02:54:04 & 1774 & 2.06
$\pm$ 0.19 &
X3.4   \\
2010-Aug-01 & 16:00:00 & W05N21 & QS & 23:18:00 & 527 & 1.22 $\pm$
0.12 &
---   \\

\tableline
\end{tabular}
\end{table}

\section{A Case Study of An Asymmetric Eruption}

Besides the fully eruptive events, partial and failed eruptions
have attracted attention in recent years. The first event of
failed eruption was studied by \citet{Ji2003}. This event can be
explained using the theory of kink instability and strong
strapping field as simulated well by \citet{Torok2005}.

A partial filament eruption on 2000 September 12, accompanied by
the M1.0 flare and CME, is also of great interest.
\citet{Wang2003} analyzed this event with multi-wavelength
observations taken by the GHN, the Extreme ultraviolet Imaging
Telescope (EIT) and LASCO. The authors tracked erupting filament
to about 20~solar radii. In the present study, we revisit this
event with a focus on the spatial distribution of decay index
along the PIL. Using the line-of-sight magnetogram from MDI, we
identified the PIL underlying the H$\alpha$ filament before the
eruption as shown in the left panel of Figure~\ref{c1}. The small
shift between the filament and the PIL is due to the projection
effect. The PIL is color-coded according to the values of decay
index, and the mean decay index over the entire PIL is about 1.64.
The white circle marks the non-erupting part of the filament. A
visual inspection immediately reveals that the non-erupting part
generally has smaller decay index than the erupting part. Panel
(b) shows some samples of potential field lines superimposed on
the line-of-sight MDI magnetogram and six given positions along
the PIL. We then quantitatively examine the values of decay index
at these six positions. The decay indexes of points 1 and 2 are
1.38 and 1.64, respectively, and the segment of the filament in
between did not erupt \citep{Wang2003}.  As a comparison, the
erupted portion of the filament lying above points 3 to 6 shows
higher decay indexes, i.e., exceeding 1.7 and up to 1.9 when the
altitude is higher than 42 Mm. The result is consistent with the
theoretical expectation that a slow decay of strapping field
contributes more to confining the eruption while a quick decay of
strapping field puts less constraint on the eruption.

\section{Discussion}

\begin{figure}
\centering
\includegraphics[scale=0.88]{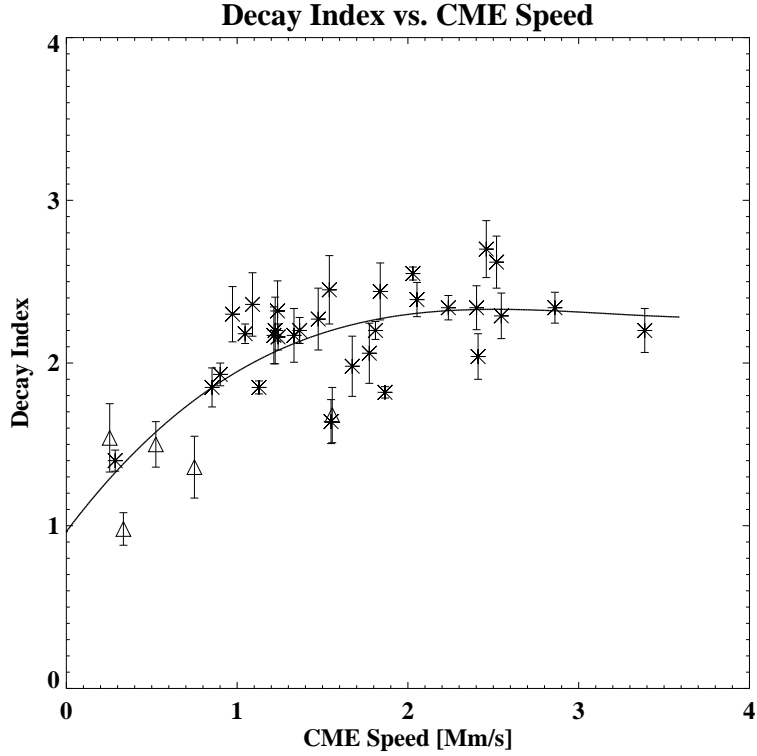}
\caption{CME speed vs. the decay index for 38 CMEs associated with
filament eruptions or two-ribbon flares events. All of the decay
indexes are calculated within an height range of 42-105 Mm and
averaged along the PIL, which are identified with the location of
filaments or flare ribbons. The decay indexes for events occurred
in the quiet Sun region are plotted with triangle symbol and all
others are plotted with Asterisks. Using a 3-degree polynomial
fit, an obvious trend is found which shows that the high speed
CMEs correspond to higher decay index in strapping fields.
\label{f1}}
\end{figure}

\begin{figure}
\centering
\includegraphics[scale=0.8]{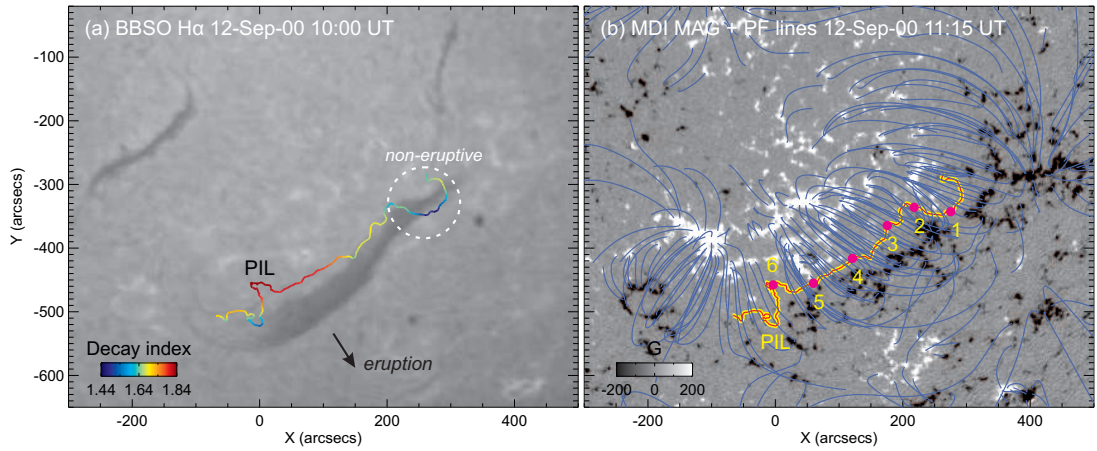}
\caption{(a) Pre-eruption H$\alpha$ image obtained from GHN,
overplotted with the spatial distribution of the decay index along
the PIL. The displacement between the filament and PIL is due to
projection effect. This filament erupted partially. The white
circle indicates the portion that did not erupt. (b) A
corresponding line-of-sight MDI magnetogram taken half hour before
the eruption, overplotted with the extrapolated potential field
lines (blue). \label{c1}}
\end{figure}

In this paper, we have analyzed 38 events that occurred not too
close to the solar limb. The decay index of each event was
calculated using the potential fields extrapolated from
line-of-sight magnetograms. Combined with the linear velocities of
CMEs obtained from the LASCO CME catalog, a good correlation is
found. We noticed that the decay index increases monotonically
with the CME speed up to 1000 km/s. For faster CMEs, the decay
index tends to have no significant change. The upper limit of
decay index is around 2.2 found by a polynomial fit to the data
points. Our observational results are consistent with the modeling
of \citet{Torok2007}. In principle, the dynamics of an eruption
may be affected by an initiation force ($F_{i}$) to accelerate the
erupting materials and the confined force ($F_{s}$) due to the
strapping field. Consequently, the kinetic energy of a CME is:
\begin{equation}
\frac{1}{2}mv^{2} = \int(F_{i} - F_{s}) dh, \label{e1}
\end{equation}

\noindent i.e.

\begin{equation}
v = \sqrt{\frac{2 \int(F_{i} - F_{s}) dh}{m}}, \label{e2}
\end{equation}

\noindent in which $m$ is mass of erupting materials, $v$ is the
CME speed and $h$ is the height. When the strapping fields decay
very fast, i.e., having large decay indexes, the $F_{s}$ is
essentially zero. From the equations above, we can see that the
CME speed is no longer correlated with decay index. This scenario
is consistent with the flattened component of the curve plotted in
Figure~\ref{f1}, in which the decay indexes are around 2.2 while
the CME speed spreads in a wide range from 1000 km/s to more than
3000 km/s. Let's consider another case that the strapping field
dominates due to a very slow decay (small decay index) in which we
have $\int(F_{s})dh \geq \int(F_{i})dh$, within a certain height
range (for instance from 42~Mm to 105~Mm). Consequently, no
filament could erupt under the constraint of such a magnetic
configuration. From previous numerical simulations,
\citep[e.g.,][]{Demoulin2010}, the lower limit of decay index
could be 1.1 for the failed eruption. Our fitted plot in
Figure~\ref{f1} indicates that at decay index equals to 1.0, CME
speed approaches to zero. Above this threshold, the speed of CME
is proportional to the square root of the net resistance force of
$\Delta F = F_{i} - F_{s}$. According to the definition, the decay
index $n = -dLog(B_{t})/dLog(h)$. The negative sign in logarithm
space indicates that $n$ is inversely proportion to the magnetic
field strength and hence the restoring force, which in our case is
$F_{s}$. We can define $F_{s} = \frac{A}{f(n)}$ and therefore
Equation~\ref{e2} becomes:

\begin{equation}
v = \sqrt{\frac{2[\int F_{i}dh - \int(\frac{A}{f(n)})dh]}{m}},
\label{e3}
\end{equation}

\noindent in which $f(n)$ is some kind of function that relates
$n$ and the force $F_{s}$, $A$ is a coefficient related to the
magnetic tension. Equation~\ref{e3} in principle resembles the
curvature component (with CME speed less than 1000 km/s) of the
plot in Figure~\ref{f1}. We note that an exact fitting to this
equation is not available due to the lack knowledge of the total
mass of CME, the magnetic tension coefficient of $A$ and the exact
form of $f(n)$. However, this equation reveals the relationship
between decay index and CME dynamics with moderate speeds.

In addition, the case study of the asymmetric distribution of
decay index of event on 2000 September 12 provides consist
physical picture that the relative high decay index indicates
weaker magnetic confinement in the strapping field. The filament
component of failed eruption was associated with lower decay index
representing a highly confined overlying arcade field. This result
is consistent with that of \citet{Liu2010a}.

Finally, we give limitations of our study. We note that only one
magnetogram prior, usually two to three hours, to each eruption is
used for extrapolation. Any rapid change of magnetic field, such
as a flux emergence is not considered during this time period
before eruptions. More detailed analysis with high cadence
magnetograms, such as observations by SDO/HMI, will provide more
accurate measurements of decay index. The results from potential
field extrapolation need to be compared with those from other
extrapolation methods, such as NLFFF. We used a height range of 42
to 105 Mm as it was commonly used in previous studies
\citep{Liu2008a, Liu2010}. But different height selections could
affect the calculation of decay index because the exact height
ranges of strapping fields vary from event to event. The filament
height, if can be measured accurately, may provide important hint
to solve this problem. In addition, a sophisticated PIL selection
tool can improve the accuracy of decay index measurements.

\acknowledgements We would like to thank the referee for valuable
comments. The CME catalog is generated and maintained by NASA and
the Catholic University of America in cooperation with the Naval
Research Laboratory. SOHO is a project of international
collaboration between ESA and NASA. The Global High Resolution
H$\alpha$ Network is supported by NSF under grant ATM-0839216.
This work is supported by NSF-AGS-0832916, NSF-AGS-0849453 and
NSF-AGS-1153424. JJ is supported by NSF grant AGS 09-36665 and
NASA grant NNX 11AQ55G.

\clearpage

\end{document}